\documentclass[final,authoryear,5p]{elsarticle}
\usepackage{epsfig}

\usepackage{amssymb}

\usepackage[ps2pdf,%
a4paper=true,%
breaklinks=true,%
colorlinks=true,%
pdfauthor={First Author et al.},%
pdftitle={Template for manuscripts in Advances in Space Research}%
]{hyperref}

\journal{Advances in Space Research}

\begin{document}

%%%%%%%%%%%%%%%%%%%%%%%%%%%%%%%%%%%%%%%%%%%%%%%%%%%%%%%%%%%%%%%%%%%%%%%%%%%%%
%% Frontmatter
\begin{frontmatter}

\title{Structural and collisional data for Mg III and Al IV}

 \author{Haykel Elabidi\corref{cor}}
 \address{Deanship of the Foundation Year, Umm Al-Qura University, Makkah, KSA;}
 \address{GRePAA, Facult\'{e} des Sciences de Bizerte,
Universit\'{e} de Carthage, Tunisia}\cortext[cor]{Corresponding
author} \ead{haelabidi@uqu.edu.sa, haykel.elabidi@fsb.rnu.tn}

\begin{abstract}
We present in this work energy levels, oscillator strengths,
radiative decay rates and fine structure collision strengths for
the Mg III and Al IV ions. The 11 configurations:(1s$^{2}$)
2s$^{2}$2p$^{6}$, 2s$^{2}$2p$^{5}3l$, 2s2p$^{6}3l$,
2s$^{2}$2p$^{5}4l$ ($l\leq n-1$, where $n$ is the principal
quantum number), yielding the lowest 75 levels are used. The
collisional data for these two ions are missing in the literature,
especially the database CHIANTI, this is the principal motivation
behind the present work. Calculations have been performed using
the AUTOSTRUCTURE code. AUTOSTRUCTURE treats the scattering
problem in the distorted wave approach. Fine structure collision
strengths are calculated for a range of electron energies from 10
Ry to 240 Ry. The atomic structure data are compared to available
experimental and theoretical results.
\end{abstract}

\begin{keyword}
%first keyword \sep second keyword \sep more keywords
plasma diagnostics; X-ray spectra; atomic structure; impact
excitation by electrons; line broadening
\end{keyword}

\end{frontmatter}

\parindent=0.5 cm

%%%%%%%%%%%%%%%%%%%%%%%%%%%%%%%%%%%%%%%%%%%%%%%%%%%%%%%%%%%%%%%%%%%%%%%%%%%%%

\section{Introduction}
Neon-like ions have a high abundance over a wide range of electron
temperatures and densities because of their closed-shell
configuration ground state. Due to its various applications in
astrophysics, plasma physics and spectroscopy, they have been the
subject of investigation for many years. The ions of this sequence
play an important role in the diagnostics of a wide variety of
laboratory and astrophysical plasmas. For example, the energies
and transition rates are used for the determination of radiative
opacities of stellar envelopes, the Opacity Project
\citep{Seaton87}, for spectral diagnostics of solar, stellar and
laboratory plasmas, for plasma modelling and for laser research,
particularly in the soft X-ray region
\citep{Lee87,Elton90,Matthews85,Feldman84}. They are used to study
transport and confinement of high-Z impurity ions in tokamaks.
Furthermore, oscillator strengths are important for the study of
laboratory and solar spectra \citep{Borges04}. Mg III and Al IV
are two ions belonging to the neon-like sequence.

The Mg III spectrum was extensively studied by \cite{Andersson71}.
Later, an experimental work on this spectrum was published by
\cite{Lundstrom73}. An extensive level classification and
wavelengths have been compiled by \citet{Kaufman91a}.
\citet{Hibbert93} have calculated configuration-interaction wave
functions in intermediate coupling for the states
2s$^{2}$2p$^{6}$, 2s$^{2}$2p$^{5}$3$l$ ($l =$ 0, 1, 2), and
2s$^{2}$2p$^{6}$3$l$ ($l =$ 0, 1, 2) of neon-like ions Ne I
through Kr XXVII, incorporating variationally optimized orbitals
and a modified Breit-Pauli Hamiltonian into the code CIV3. They
have presented the percentage $LS$ compositions of the $LSJ$
levels, together with their energies, oscillator strengths, and
probabilities of transition between them. Lifetimes of
2s$^{2}$2p$^{5}$3$l$ levels have been also presented. More
recently and using the multi-configuration Hartree-Fock (MCHF)
method with relativistic effects, \citet{Fischer04} have obtained
energy levels and transitions probabilities for transitions
between computed levels for the Be-like ($4\leq Z \leq 12$) to
Ne-like ($10\leq Z \leq 24$) sequences including the Mg III and Al
IV ions ($Z$ is the nuclear charge). A recent experimental study
of Mg III was published \citep{Brown09}. Further analyzes of the
Mg III spectra have been performed and about 60 unobserved levels
have been predicted \citep{LiangII10} using the multichannel
quantum defect theory (MQDT). The most recent results are those of
\citet{Beiersdorfer11}, where the radiative decay rates of the
(2s$^{2}$2p$_{1/2}^{5}$3s$_{1/2}$)$_{J=0}$ level in neon-like ions
have been calculated for nuclear charges ranging from Z=10 to
Z=110.

In \citet{Artru75}, a total of 225 new lines of Al IV have been
observed in the wavelength range of 400$-$4700 $\rm \AA$, leading
to the determination of all of the levels of the 2p$^{5}$4p, 4d,
4f, 5s, 5f, and 5g configurations. \citet{Martin79} have published
energy levels for the atom and all positive ions of aluminum,
where the authors have critically compiled their data using the
published material on measurements and analyzes of the optical
spectra. An extensive level classification and wavelengths for Al
IV have been compiled by \citet{Kaufman91b}. A more recent
compilation of atomic transition probabilities for about 5000
lines of aluminium in all its ionization stages (except the
hydrogenic one) has been published by \citet{Kelleher08}.

To our best knowledge, there are no distorted wave fine structure
collision strengths for the Mg III and Al IV ions to compare with.
Although the number of published papers dedicated to the atomic
structure of Mg III and Al IV is important, the only published
electron scattering calculations are those of \citet{Ganas80} and
\citet{Liang10}. In \citet{Ganas80}, integrated cross sections for
incident electron energies ranging from threshold to 5 keV were
calculated for the 2p$–$3s resonance transition along the
neon-like sequence. \textbf{The authors in \citet{Ganas80} used an
analytic atomic independent particle model potential adjusted to
experimental energy levels to generate wave functions for the
ground and excited states of the considered ions. The obtained
wave functions are used in conjunction with the Born approximation
and the $LS-$coupling scheme to obtain generalized oscillator
strengths, which are used to calculate integrated cross sections.
The method of deriving generalized oscillator strengths and the
formula used to obtain cross sections may be found in
\citet{Ganas98}.} In \citet{Liang10}, electron impact excitation
data were calculated for Ne-like ions from Na II to Kr XXVII using
the intermediate-coupling frame transformation R-matrix approach,
and the results of effective collision strengths were presented.

The principal motivation behind the present work is the missed
collisional data for the two ions Mg III and Al IV in many
databases. These data can be useful for many astrophysical
investigations. Besides the importance of the structural and
collisional data of multicharged ions in astrophysical and
laboratory plasmas investigation, collision strengths have another
importance which is related to another field of investigation: the
comparison of these data with experimental or/and other
theoretical results can be a powerful tool to check our line
broadening calculations
\citep{Elabidi08,Elabidi09,Elabidi11a,Elabidi11b,Elabidi12}.
Indeed, our line broadening method is ab initio, this means that
all the parameters required for the line broadening calculations
such as radiative atomic data (energy levels, oscillator
strengths...) and collisional data (collision strengths or cross
sections, scattering matrices...) are evaluated during the
calculation and not taken from other data sources. Consequently,
the accuracy of our broadening parameters is strongly related to
the accuracy of the atomic and collisional ones. This represents
an other motivation and interest of the present work.

The aim of this paper is to provide fine structure collision
strengths for Mg III and Al IV transitions in the distorted wave
approximation. The atomic structure has been calculated for the 75
levels arising from the eleven configurations ($1s^{2}$)
$2s^{2}2p^{6}$, $2s^{2}2p^{5}3l$, $2s2p^{6}3l$, $2s^{2}2p^{5}4l$
($l\leq n-1$). Collision strengths have been computed for
transitions from the ground and the four first excited levels to
all the levels. The incoming electron energies used in our
calculations are between 10 Ry and 240 Ry. Discussions and
investigations of convergence of collision strengths with energy
and with total angular momentum $J^{T}$ are also given. Only the
atomic structure data are compared to available experimental and
theoretical results.

\section{Atomic structure and electron-ion scattering}
The atomic structure has been calculated using the AUTOSTRUCTURE
(AS) code \citep{Badnell86,Badnell97} by constructing target
wavefunctions using radial wavefunctions calculated in a scaled
Thomas-Fermi-Dirac-Amaldi statistical model potential using the
Breit-Pauli intermediate coupling \citep{Bethe57}. The radial
scaling parameters $\lambda_{nl}$ (depending on $n$ and $l$) are
determined by minimizing the sum of the energies of all the target
terms, computed in $LS$ coupling, i.e. neglecting all relativistic
effects. In this code, besides the one-body and the two-body fine
structure interactions, the two-body non-fine structure operators
of the Breit-Pauli Hamiltonian, namely contact spin-spin, two-body
Darwin and orbit-orbit are incorporated. More details of how these
interactions are incorporated are reported in \citet{Badnell97}.

Recently, the Breit-Pauli Distorted Wave (BPDW) approach for
electron impact excitation of atomic ions has been implemented in
the AS code \citep{Badnell11}, which we use for the scattering
problem in the present paper. We note that the distorted wave
approximation (DW) is adequate for moderately and highly charged
ions and the agreement between the DW and more sophisticated
methods (close coupling for example) is good. Collision strengths
are calculated at the same set of final scattered energies for all
transitions: zero gives all threshold transitions, for example.
For large $l$ values, a 'top up' for dipole transitions makes use
of the sum rule of \citet{Burgess74}. For higher multipoles, a
geometric series in energy in combination with the degenerate
energy limit \citep{Burgess70} is used to take into account of
large $l$ contributions to collision strengths.

\section{Results and discussions}
\subsection{Structure}

Eleven configurations: ($1s^{2}$) $2s^{2}2p^{6}$,
$2s^{2}2p^{5}3l$, $2s2p^{6}3l$, $2s^{2}2p^{5}4l$ ($l\leq n-1$)
have been used in AUTOSTRUCTURE to study the atomic structure of
the Mg III and Al IV ions. This set of configurations gives rise
to 75 fine structure levels. The radial scaling parameters
$\lambda_{nl}$ used in the code AUTOSTRUCTURE for the two ions are
listed in the Table \ref{tab0}. The energy levels of Mg III are
listed in Table \ref{tab1}. We have presented a comparison of our
Mg III energies with the NIST \citep{NIST12} values, with those of
\citet{Fischer04} and with the results of \citet{Liang10}. The
wavefunctions in \citet{Fischer04} were determined using the
multi-configuration Hartree–Fock (MCHF) method with relativistic
effects included through the Breit–Pauli Hamiltonian, where only
the orbit-orbit interaction was omitted. \textbf{In
\citet{Liang10}, the authors used the code AUTOSTRUCTURE but with
a 31-configuration model.} Our ground level $1s^{2}2s^{2}2p^{6}$
$^{1}$S$_{0}$ energy has been shifted by +24993 cm$^{-1}$. This
shift is obtained by the difference between the center of gravity
of levels 2$-$27 for calculated and compiled energies by NIST
\citep{NIST12}. After adjustment, our results become in an
excellent agreement (less than 1 \%) with those of
\citet{Fischer04} and \citet{Liang10}. \textbf{It is important to
note that, even the authors in \citet{Liang10} used more
configurations than we used in the present work, we prefer to
present our energy levels and compare them with other results.
This is because the collision strengths presented in the next
subsection are calculated using our 11-configuration model
energies. So, it is better to present coherent structural and
collisional data derived from the same model.} We present in Table
\ref{tab2} our line strengths, oscillator strengths and transition
probabilities for spontaneous emission for the Mg III allowed
transitions (E1). The comparison between our results and those of
the MCHF calculations shows that the relative differences between
them are about 12 \% for line strengths, 11 \% for oscillator
strengths and 10 \% for transition probabilities. Our Al IV
adjusted energies are also compared with the MCHF
\citep{Fischer04} results and with those of \citet{Liang10} in
Table \ref{tab3}. The same procedure of the ground level
adjustment has been adopted as for Mg III, and we have shifted the
ground level energy by +22339 cm$^{-1}$. Excellent agreement has
been found between our results and the two other ones. Line
strengths, oscillator strengths and transition probabilities of Al
IV are presented in Table \ref{tab4}. The relative differences
between our results and those of \citet{Fischer04} are of the same
order of magnitude as those of the Mg III ion. The averaged
relative difference between the two methods is about 10 \%. In the
following, some common remarks of the two ions will be drawn.
Firstly, an inversion of some levels is found between our energies
and those of NIST \citep{NIST12} and \citet{Fischer04}. The
inverted levels are denoted by asterisks in Tables \ref{tab1} and
\ref{tab3}. Secondly, our adjusted energies have been used
(instead of the calculated ones) to evaluate the present radiative
data (line strengths, oscillator strengths and transition
probabilities). Thirdly, the worse disagreement for level energies
is about 1.6 \% and has been found for the level $2s^{2}2p^{5}3p$
$^{1}$S$_{0}$ (level 15 in Tables \ref{tab1} and \ref{tab3}). We
found that the energy of level 15 before the adjusting process had
the best agreement with the MCHF results of \citet{Fischer04}
(about 1.6 \%), and all the other energies have a relative
difference of about 3 \%. After adjustment of our ground level,
all the level energies become in agreement with the MCHF results
except the energy of the level 15. Finally, we found that the
highest difference (about 30 \%) between our results and those of
the MCHF method \citep{Fischer04} has been found for the two
transitions $2s^{2}2p^{5}3s$
$^{1}$P$_{1}^{\mathrm{o}}-2s^{2}2p^{5}3p$ $^{3}$D$_{2}$ and
$2s^{2}2p^{5}3s$ $^{1}$P$_{1}^{\mathrm{o}}-2s^{2}2p^{5}3p$
$^{3}$D$_{1}$ (transitions $5-8$ and $5-9$ in Tables \ref{tab2}
and \ref{tab4}).

\subsection{Collision problem}
Collision problem has been treated in the distorted wave
approximation using the AUTOSTRUCTURE code. We present in Tables
\ref{tab5}, \ref{tab6}, \ref{tab7}, \ref{tab8} and \ref{tab9} Mg
III collision strengths from the ground level and from the first
four excited levels to all the other levels. Our collision
strengths have been computed for five energies 10, 20, 40, 80 and
160 Ry. In Tables \ref{tab10}, \ref{tab11}, \ref{tab12},
\ref{tab13} and \ref{tab14} Al IV collision strengths from the
ground level and from the first four excited levels to all the
other levels are presented. The incoming electron energies used
are 15, 30, 60, 120 and 240 Ry. To our best knowledge, there are
no distorted wave collision strengths for Mg III or Al IV to
compare with. In \citet{Liang10}, electron impact excitation data
were calculated for Ne-like ions from Na II to Kr XXVII using the
intermediate-coupling frame transformation R-matrix approach, and
only the results of effective collision strengths were presented.
Any other collision calculations for these two ions will be very
helpful for two reasons. Firstly, to compare with our results and
to decide about their applicability in astrophysics and plasma
diagnostics. Secondly, since collision strengths (and other
collisional parameters like scattering matrices, cross
sections...) are used in our line broadening calculations, their
comparison with collision strengths derived from other different
methods will be very interesting for the evaluation of our line
broadening results. \textbf{We hope that the structural and
collisional data presented in this paper will be useful in
spectral diagnostics and modelling of astrophysical and laboratory
plasmas, laser development and tokamaks research.}

\section{Conclusion}
We have calculated energy levels, oscillator strengths and
radiative decay rates for the two neon-like ions Mg III and Al IV.
We have used eleven configurations: ($1s^{2}$) $2s^{2}2p^{6}$,
$2s^{2}2p^{5}3l$, $2s2p^{6}3l$, $2s^{2}2p^{5}4l$ ($l\leq n-1$)
yielding 75 fine structure levels. The atomic structure has been
studied using the AUTOSTRUCTURE code. We have compared our level
energies with the NIST values and with the multiconfiguration
Hartree-Fock ones. We find that the agreement (after adjustment of
the ground level energy) is much better than 1 \%. The agreement
for the oscillator strengths and the radiative decay rates is
about 12 \%.

Fine structure collision strengths have been calculated in the
distorted wave approximation using the AUTOSTRUCTURE code at five
electron energies: 10, 20, 40, 80 and 160 Ry for Mg III and 15,
30, 60, 120 and 240 Ry for Al IV. We have presented our collision
strengths between the ground and the four first excited levels to
all other ones. We have taken into account the importance of high
partial-wave contribution to collision strengths.

\section*{Acknowledgments}
This work has been supported by the Tunisian research unit
05/UR/12-04.

\newpage

\begin{table}
\caption{Radial scaling parameters $\lambda_{nl}$ used in
AUTOSTRUCTURE.}
\begin{tabular}{ccccccccccc}
\hline
Ion & 1s & 2s & 2p & 3s & 3p & 3d & 4s & 4p & 4d & 4f \\
Mg & 1.6229 & 1.1129 & 1.0581 & 0.9879 & 0.9582 & 0.9140 & 1.0035 & 0.9466 & 0.9115 & 0.8629 \\
Al & 1.6596 & 1.1231 & 1.0641 & 1.0068 & 0.9709 & 0.9411 & 1.0243 & 0.9599 & 0.9401 & 0.9034 \\
\hline \label{tab0}
\end{tabular}
\end{table}

\begin{table}
\caption{Mg III energy levels in cm$^{-1}$. $E$: present results,
$E_{\rm NIST}$: energies reported by NIST \citep{NIST12} and taken
from \citet{Martin80}, $E_{\rm MCHF}$: Hartree-Fock energies
\citep{Fischer04}, $E_{\rm LB10}$: energies calculated in
\citet{Liang10} by AUTOSTRUCTURE using 31 configurations. Levels
denoted by asterisks ($*$) are inverted compared to our values.}
\begin{tabular}{ccccccc}
\hline
$i$ & Conf. & Level & $E$ & $E_{\rm NIST}$ & $E_{\rm MCHF}$&$E_{\rm LB10}$ \\
\hline
$1$ & $2s^{2}2p^{6}$ & $^{1}$S$_{0}$ & $0$ & $0$ & $0$&0 \\
$2$ & $2s^{2}2p^{5}3s$ & $^{3}$P$_{2}^{\mathrm{o}}$ & 426260 & 425640.3 &425638.16&424178.0 \\
$3$ & $2s^{2}2p^{5}3s$ & $^{3}$P$_{1}^{\mathrm{o}}$ & 427481 & 426868.1 &426861.97&425391.0 \\
$4$ & $2s^{2}2p^{5}3s$ & $^{3}$P$_{0}^{\mathrm{o}}$ & 428462 & 427852.1 &427834.77&426313.0 \\
$5$ & $2s^{2}2p^{5}3s$ & $^{1}$P$_{1}^{\mathrm{o}}$ & 432100 & 431530.0 &431569.27&430399.0 \\
$6$ & $2s^{2}2p^{5}3p$ & $^{3}$S$_{1}$ & 468025 & 467378.5 & 467378.97&465349.0 \\
$7$ & $2s^{2}2p^{5}3p$ & $^{3}$D$_{3}$ & 474034 & 474053.2 & 474052.45&472851.0 \\
\end{tabular}
\end{table}

\clearpage

\addtocounter{table}{-1}
\begin{table}
\caption{Continued.}
\begin{tabular}{ccccccc}
\hline
$i$ & Conf. & Level & $E$ & $E_{\rm NIST}$ & $E_{\rm MCHF}$&$E_{\rm LB10}$ \\
\hline
$43$ & $2s^{2}2p^{5}4d$ & $^{3}$P$_{1}^{\mathrm{o}}$ & 580127 &  &&580886.0  \\
$44$ & $2s^{2}2p^{5}4d$ & $^{3}$P$_{2}^{\mathrm{o}}$ & 580457 &  &&581443.0  \\
$45$ & $2s^{2}2p^{5}4d$ & $^{3}$F$_{4}^{\mathrm{o}}$ & 580476 &  &&581590.0  \\
$46$ & $2s^{2}2p^{5}4d$ & $^{3}$F$_{3}^{\mathrm{o}}$ & 580662 &  &&581800.0  \\
$47$ & $2s^{2}2p^{5}4d$ & $^{1}$D$_{2}^{\mathrm{o}}$ & 581092 &  &&582275.0  \\
$48$ & $2s^{2}2p^{5}4d$ & $^{3}$D$_{3}$ & 581187 &  &&582408.0  \\
$49$ & $2s^{2}2p^{5}4d$ & $^{3}$D$_{1}^{\mathrm{o}}$ & 581937 &  &&583236.0  \\
\hline
\end{tabular}
\end{table}

\clearpage

\begin{table}
\caption{Line strengths ($S$), oscillator strengths ($f_{ij}$),
and radiative decay rates ($A_{ji}$) for some Mg III lines. Our
results (Present) are compared with the Hartree-Fock (MCHF) values
\citep{Fischer04}.}
\begin{tabular}{ccccccc}
\hline
Transition & \multicolumn{2}{c}{$S$} & \multicolumn{2}{c}{$f_{ij}$} & \multicolumn{2}{c}{$A_{ji}$ ($s^{-1}$)} \\
\hline
$i-j$ & Present & MCHF & Present & MCHF & Present & MCHF \\
\hline
$1-3$ & 1.228E$-$02 & 8.946E$-$03 & 1.594E-02 & 1.160E$-$02 & 6.477E+08 &4.700E$+$08 \\
$1-5$ & 2.113E$-$01 & 1.687E$-$01 & 2.773E-01 & 2.212E$-$01 & 1.151E+10 &9.160E$+$09 \\
$1-17$ & 2.424E$-$03 & 1.931E$-$03 & 3.897E-03 & 3.111E$-$03 & 2.428E+08 &1.946E$+$08 \\
$1-23$ & 1.367E$-$01 & 1.220E$-$01 & 2.215E-01 & 1.980E$-$01 & 1.401E+10 &1.257E$+$10 \\
$2-6$ & 3.792E$+$00 & 3.404E$+$00 & 9.621E$-$02 & 8.631E$-$02 &1.866E$+$08& 1.672E$+$08 \\
$2-7$ & 1.390E$+$01 & 1.282E$+$01 & 4.035E$-$01 & 3.772E$-$01 &4.388E$+$08& 4.212E$+$08 \\
\end{tabular}
\end{table}

\clearpage

\begin{table}
\caption{Al IV energy levels in cm$^{-1}$. $E$: present results,
$E_{\rm NIST}$: energies reported by NIST \citep{NIST12} and taken
from \citet{Martin79}, $E_{\rm MCHF}$: Hartree-Fock energies
\citep{Fischer04}, $E_{\rm LB10}$: energies calculated in
\citet{Liang10} by AUTOSTRUCTURE using 31 configurations. Levels
denoted by asterisks ($*$) are inverted compared to our values.}
\begin{tabular}{ccccccc}
\hline
$i$ & Conf. & Level & $E$ & $E_{\rm NIST}$ & $E_{\rm MCHF}$&$E_{\rm LB10}$ \\
\hline
$1$ & $2s^{2}2p^{6}$ & $^{1}$S$_{0}$ & $0$ & $0$ & 0&0 \\
$2$ & $2s^{2}2p^{5}3s$ & $^{3}$P$_{2}^{\mathrm{o}}$ & 616706 &616644.2 &616643.56&615386.0 \\
$3$ & $2s^{2}2p^{5}3s$ & $^{3}$P$_{1}^{\mathrm{o}}$ & 618536 &618473.9 &618470.83&617213.0 \\
$4$ & $2s^{2}2p^{5}3s$ & $^{3}$P$_{0}^{\mathrm{o}}$ & 620113 &620060.1 &620050.03&618711.0 \\
$5$ & $2s^{2}2p^{5}3s$ & $^{1}$P$_{1}^{\mathrm{o}}$ & 624841 &624717.5 &624721.95&623896.0 \\
$6$ & $2s^{2}2p^{5}3p$ & $^{3}$S$_{1}$ & 671875 & 671632.5 & 671634.22&669906.0 \\
$7$ & $2s^{2}2p^{5}3p$ & $^{3}$D$_{3}$ & 680515 & 680859.8 & 680859.51&679917.0 \\
\end{tabular}
\end{table}

\clearpage

\addtocounter {table}{-1}
\begin{table}
\caption{Continued.}
\begin{tabular}{ccccccc}
\hline
$i$ & Conf. & Level & $E$ & $E_{\rm NIST}$ & $E_{\rm MCHF}$&$E_{\rm LB10}$ \\
\hline
$43$ & $2s^{2}2p^{5}4d$ & $^{3}$P$_{1}^{\mathrm{o}}$ & 849921 &  &&852068.0  \\
$44$ & $2s^{2}2p^{5}4d$ & $^{3}$P$_{2}^{\mathrm{o}}$ & 850465 &  &&852670.0  \\
$45$ & $2s^{2}2p^{5}4d$ & $^{3}$F$_{4}^{\mathrm{o}}$ & 850529 &  &&852914.0  \\
$46$ & $2s^{2}2p^{5}4d$ & $^{3}$F$_{3}^{\mathrm{o}}$ & 850896 &  &&853306.0  \\
$47$ & $2s^{2}2p^{5}4d$ & $^{3}$D$_{3}$ & 851575 &  &&854328.0  \\
$48$ & $2s^{2}2p^{5}4d$ & $^{1}$D$_{2}$ & 851804 &  &&854036.0  \\
$49$ & $2s^{2}2p^{5}4d$ & $^{3}$D$_{1}$ & 853286 &  &&855841.0  \\
\hline
\end{tabular}
\end{table}

\clearpage

\begin{table}
\caption{Line strengths ($S$), oscillator strengths ($f_{ij}$),
and radiative decay rates ($A_{ji}$) for some Al IV lines. Our
results (Present) are compared with the Hartree-Fock (MCHF) values
\citep{Fischer04}.}
\begin{tabular}{ccccccc}
\hline
Transition & \multicolumn{2}{c}{$S$} & \multicolumn{2}{c}{$f_{ij}$} & \multicolumn{2}{c}{$A_{ji}$($s^{-1}$)} \\
\hline
$i-j$ & Present & MCHF & Present & MCHF &Present & MCHF \\
\hline
$1-3$ & 1.075E-02 & 9.144E$-$03 & 2.020E-02& 1.718E$-$02 & 1.719E+09 &1.461E$+$09 \\
$1-5$ & 1.391E-01 & 1.177E$-$01 & 2.640E-01 & 2.234E$-$01 &2.291E+10 &1.939E$+$10 \\
$1-17$ & 1.374E-03 & 1.264E$-$03 & 3.165E-03 & 2.916E$-$03 &4.047E+08 &3.741E$+$08 \\
$1-23$ & 5.067E-02 & 5.796E$-$02 & 1.179E-01 & 1.350E$-$01 &1.537E+10 &1.765E$+$10 \\
$2-6$ & 2.546E+00 & 2.322E+00 & 8.533E-02 & 7.756E-02 & 2.887E+08& 2.607E+08\\
$2-7$ & 9.417E+00 & 8.741E+00 & 3.651E-01 & 3.410E-01 & 7.081E+08& 6.700E+08\\
\end{tabular}
\end{table}

\clearpage

\begin{table}
\caption{Fine structure collision strengths from the Mg III ground
level $1s^{2}2s^{2}2p^{6}$ $^{1}$S$_{0}$ (level 1 in Table
\ref{tab1}) to all the other levels.}
\begin{tabular}{cccccc}
\hline
$i-j$ & $10$ Ry & $20$ Ry & $40$ Ry & $80$ Ry & $160$ Ry \\
\hline
$1-2$ & 3.120E-03 & 1.673E-03 & 6.358E-04 & 1.868E-04 & 5.347E-05 \\
$1-3$ & 2.491E-02 & 3.138E-02 & 3.338E-02 & 3.921E-02 & 4.982E-02 \\
$1-4$ & 6.223E-04 & 3.338E-04 & 1.269E-04 & 3.726E-05 & 1.065E-05 \\
$1-5$ & 3.943E-01 & 5.218E-01 & 5.662E-01 & 6.694E-01 & 8.523E-01 \\
$1-6$ & 9.857E-03 & 4.755E-03 & 1.867E-03 & 6.209E-04 & 1.978E-04 \\
$1-7$ & 6.458E-03 & 3.060E-03 & 1.137E-03 & 3.450E-04 & 9.508E-05 \\
$1-8$ & 8.165E-03 & 8.644E-03 & 9.572E-03 & 1.061E-02 & 8.867E-03 \\
\end{tabular}
\end{table}

\clearpage

\addtocounter{table}{-1}
\begin{table}
\caption{Continued.}
\begin{tabular}{cccccc}
\hline
$i-j$ & $10$ Ry & $20$ Ry & $40$ Ry & $80$ Ry & $160$ Ry \\
\hline
$1-42$ & 3.563E-04 & 1.137E-04 & 3.159E-05 & 7.239E-06 & 1.436E-06 \\
$1-43$ & 4.425E-03 & 5.584E-03 & 5.840E-03 & 7.301E-03 & 1.096E-02 \\
$1-44$ & 1.449E-03 & 4.561E-04 & 1.251E-04 & 2.861E-05 & 5.679E-06 \\
$1-45$ & 1.151E-03 & 3.393E-04 & 8.877E-05 & 2.027E-05 & 4.139E-06 \\
$1-46$ & 2.133E-03 & 2.553E-03 & 3.331E-03 & 3.822E-03 & 3.844E-03 \\
$1-47$ & 4.644E-04 & 1.252E-04 & 2.972E-05 & 6.733E-06 & 1.419E-06\\
$1-48$ & 1.657E-03 & 2.043E-03 & 2.707E-03 & 3.123E-03 & 3.148E-03 \\
\hline
\end{tabular}
\end{table}

\clearpage

\begin{table}
\caption{Fine structure collision strengths from the Mg III level
$1s^{2}2s^{2}2p^{5}3s$ $^{3}$P$^{\rm o}_{2}$ (level 2 in Table
\ref{tab1}) to all the other levels.}
\begin{tabular}{cccccc}
\hline
$i-j$ & $10$ Ry & $20$ Ry & $40$ Ry & $80$ Ry & $160$ Ry \\
\hline
$2-3$ & 2.095E-01 & 1.527E-01 & 1.339E-01 & 1.328E-01 & 1.366E-01 \\
$2-4$ & 6.579E-02 & 5.879E-02 & 5.861E-02 & 6.109E-02 & 6.375E-02 \\
$2-5$ & 1.519E-01 & 7.291E-02 & 3.097E-02 & 1.466E-02 & 9.603E-03 \\
$2-6$ & 2.047E+01 & 2.450E+01 & 2.863E+01 & 3.267E+01 & 3.714E+01 \\
$2-7$ & 7.013E+01 & 8.451E+01 & 9.973E+01 & 1.147E+02 & 1.313E+02 \\
$2-8$ & 1.770E+01 & 2.126E+01 & 2.506E+01 & 2.882E+01 & 3.295E+01 \\
$2-9$ & 2.277E+00 & 2.733E+00 & 3.224E+00 & 3.711E+00 & 4.245E+00 \\
\end{tabular}
\end{table}

\clearpage

\addtocounter {table}{-1}
\begin{table}
\caption{continued.}
\begin{tabular}{cccccc}
\hline
$i-j$ & $10$ Ry & $20$ Ry & $40$ Ry & $80$ Ry & $160$ Ry \\
\hline
$2-42$ & 2.461E-02 & 2.961E-02 & 3.327E-02 & 4.031E-02 & 3.755E-02 \\
$2-43$ & 6.854E-02 & 8.239E-02 & 9.270E-02 & 1.122E-01 & 1.038E-01 \\
$2-44$ & 8.797E-02 & 1.051E-01 & 1.185E-01 & 1.422E-01 & 1.293E-01 \\
$2-45$ & 2.313E-01 & 2.786E-01 & 3.162E-01 & 3.809E-01 & 3.424E-01 \\
$2-46$ & 6.226E-02 & 7.289E-02 & 8.204E-02 & 9.890E-02 & 8.895E-02 \\
$2-47$ & 4.168E-02 & 4.867E-02 & 5.480E-02 & 6.576E-02 & 5.890E-02 \\
$2-48$ & 1.289E-01 & 1.543E-01 & 1.751E-01 & 2.099E-01 & 1.865E-01 \\
\hline
\end{tabular}
\end{table}

\clearpage

\begin{table}
\caption{Fine structure collision strengths from the Mg III level
$1s^{2}2s^{2}2p^{5}3s$ $^{3}$P$^{\rm o}_{1}$ (level 3 in Table
\ref{tab1}) to all the other levels.}
\begin{tabular}{cccccc}
\hline
$i-j$ & $10$ Ry & $20$ Ry & $40$ Ry & $80$ Ry & $160$ Ry \\
\hline
$3-4$ & 5.843E-02 & 2.442E-02 & 8.466E-03 & 2.544E-03 & 6.812E-04 \\
$3-5$ & 8.958E-02 & 4.994E-02 & 3.075E-02 & 2.430E-02 & 2.291E-02 \\
$3-6$ & 9.343E+00 & 1.116E+01 & 1.301E+01 & 1.482E+01 & 1.681E+01 \\
$3-7$ & 8.263E-03 & 2.147E-03 & 6.073E-04 & 3.764E-04 & 2.470E-04 \\
$3-8$ & 3.209E+01 & 3.860E+01 & 4.548E+01 & 5.225E+01 & 5.968E+01 \\
$3-9$ & 1.689E+01 & 2.031E+01 & 2.393E+01 & 2.751E+01 & 3.142E+01 \\
$3-10$ & 5.261E+00 & 6.322E+00 & 7.469E+00 & 8.609E+00 & 9.861E+00 \\
\end{tabular}
\end{table}

\clearpage

\addtocounter {table}{-1}
\begin{table}
\caption{continued.}
\begin{tabular}{cccccc}
\hline
$i-j$ & $10$ Ry & $20$ Ry & $40$ Ry & $80$ Ry & $160$ Ry \\
\hline
$3-42$ & 1.512E-04 & 3.897E-05 & 6.479E-06 & 9.360E-07 & 1.360E-07 \\
$3-43$ & 8.665E-03 & 9.971E-03 & 1.106E-02 & 1.276E-02 & 1.177E-02 \\
$3-44$ & 3.493E-02 & 4.184E-02 & 4.691E-02 & 5.708E-02 & 5.314E-02 \\
$3-45$ & 1.445E-03 & 3.822E-04 & 7.190E-05 & 1.846E-05 & 8.076E-06 \\
$3-46$ & 8.075E-02 & 9.751E-02 & 1.102E-01 & 1.337E-01 & 1.217E-01 \\
$3-47$ & 6.463E-02 & 7.806E-02 & 8.831E-02 & 1.069E-01 & 9.682E-02 \\
$3-48$ & 2.553E-02 & 3.000E-02 & 3.398E-02 & 4.069E-02 & 3.591E-02 \\
\hline
\end{tabular}
\end{table}

\clearpage

\begin{table}
\caption{Fine structure collision strengths from the Mg III level
$1s^{2}2s^{2}2p^{5}3s$ $^{3}$P$^{\rm o}_{0}$ (level 4 in Table
\ref{tab1}) to all the other levels.}
\begin{tabular}{cccccc}
\hline
$i-j$ & $10$ Ry & $20$ Ry & $40$ Ry & $80$ Ry & $160$ Ry \\
\hline
$4-5$ & 2.749E-02 & 1.186E-02 & 4.098E-03 & 1.207E-03 & 3.153E-04 \\
$4-6$ & 2.757E+00 & 3.290E+00 & 3.828E+00 & 4.354E+00 & 4.936E+00 \\
$4-7$ & 1.776E-04 & 2.006E-04 & 2.119E-04 & 1.921E-04 & 1.310E-04 \\
$4-8$ & 1.305E-03 & 3.039E-04 & 4.235E-05 & 5.729E-06 & 1.259E-06 \\
$4-9$ & 1.102E+01 & 1.324E+01 & 1.559E+01 & 1.789E+01 & 2.042E+01 \\
$4-10$ & 7.658E-04 & 1.679E-04 & 1.873E-05 & 1.257E-06 & 1.633E-07 \\
$4-11$ & 8.847E+00 & 1.064E+01 & 1.257E+01 & 1.448E+01 & 1.656E+01 \\
\end{tabular}
\end{table}

\clearpage

\addtocounter {table}{-1}
\begin{table}
\caption{continued.}
\begin{tabular}{cccccc}
\hline
$i-j$ & $10$ Ry & $20$ Ry & $40$ Ry & $80$ Ry & $160$ Ry \\
\hline
$4-42$ & 8.822E-04 & 9.701E-04 & 9.609E-04 & 8.474E-04 & 6.138E-04 \\
$4-43$ & 1.610E-05 & 4.440E-06 & 8.372E-07 & 1.424E-07 & 2.531E-08 \\
$4-44$ & 3.127E-03 & 3.761E-03 & 4.170E-03 & 5.090E-03 & 4.993E-03 \\
$4-45$ & 1.058E-05 & 8.789E-06 & 6.314E-06 & 5.568E-06 & 3.928E-06 \\
$4-46$ & 1.582E-05 & 4.725E-06 & 1.093E-06 & 2.391E-07 & 5.200E-08 \\
$4-47$ & 2.492E-03 & 3.018E-03 & 3.367E-03 & 4.130E-03 & 3.950E-03 \\
$4-48$ & 2.375E-05 & 6.046E-06 & 9.989E-07 & 1.433E-07 & 2.064E-08 \\
\hline
\end{tabular}
\end{table}

\clearpage

\begin{table}
\caption{Fine structure collision strengths from the Mg III level
$1s^{2}2s^{2}2p^{5}3s$ $^{1}$P$^{\rm o}_{1}$ (level 5 in Table
\ref{tab1}) to all the other levels.}
\begin{tabular}{cccccc}
\hline
$i-j$ & $10$ Ry & $20$ Ry & $40$ Ry & $80$ Ry & $160$ Ry \\
\hline
$5-6$ & 2.010E-01 & 2.350E-01 & 2.709E-01 & 3.064E-01 & 3.460E-01 \\
$5-7$ & 6.024E-03 & 1.322E-03 & 1.626E-04 & 2.918E-05 & 1.558E-05 \\
$5-8$ & 9.277E-01 & 1.106E+00 & 1.293E+00 & 1.481E+00 & 1.690E+00 \\
$5-9$ & 1.761E-01 & 2.062E-01 & 2.406E-01 & 2.758E-01 & 3.148E-01 \\
$5-10$ & 2.891E+01 & 3.474E+01 & 4.083E+01 & 4.685E+01 & 5.350E+01 \\
$5-11$ & 1.689E+01 & 2.028E+01 & 2.385E+01 & 2.738E+01 & 3.126E+01 \\
$5-12$ & 2.296E+01 & 2.759E+01 & 3.248E+01 & 3.732E+01 & 4.264E+01 \\
\end{tabular}
\end{table}

\clearpage

\addtocounter {table}{-1}
\begin{table}
\caption{continued.}
\begin{tabular}{cccccc}
\hline
$i-j$ & $10$ Ry & $20$ Ry & $40$ Ry & $80$ Ry & $160$ Ry \\
\hline
$5-42$ & 1.230E-04 & 3.196E-05 & 5.519E-06 & 8.425E-07 & 1.314E-07 \\
$5-43$ & 8.574E-04 & 6.493E-04 & 6.411E-04 & 6.545E-04 & 5.269E-04 \\
$5-44$ & 5.765E-03 & 6.284E-03 & 6.909E-03 & 8.413E-03 & 7.388E-03 \\
$5-45$ & 1.063E-03 & 2.716E-04 & 4.481E-05 & 6.906E-06 & 1.342E-06 \\
$5-46$ & 3.994E-02 & 4.757E-02 & 5.312E-02 & 6.529E-02 & 5.913E-02 \\
$5-47$ & 2.785E-02 & 3.300E-02 & 3.689E-02 & 4.522E-02 & 4.060E-02 \\
$5-48$ & 4.187E-02 & 4.996E-02 & 5.581E-02 & 6.863E-02 & 6.258E-02 \\
\hline
\end{tabular}
\end{table}

\clearpage

\begin{table}
\caption{Fine structure collision strengths from the Al IV ground
level $1s^{2}2s^{2}2p^{6}$ $^{1}$S$_{0}$ (level 1 in Table
\ref{tab3}) to all the other levels.}
\begin{tabular}{cccccc}
\hline
$i-j$ & $15$ Ry & $30$ Ry & $60$ Ry & $8120$ Ry & $240$ Ry \\
\hline
$1-2$ & 2.480E-03 & 1.217E-03 & 4.371E-04 & 1.254E-04 & 3.629E-05 \\
$1-3$ & 2.146E-02 & 2.696E-02 & 2.890E-02 & 3.420E-02 & 4.390E-02 \\
$1-4$ & 4.946E-04 & 2.428E-04 & 8.716E-05 & 2.497E-05 & 7.211E-06 \\
$1-5$ & 2.574E-01 & 3.390E-01 & 3.694E-01 & 4.391E-01 & 5.646E-01 \\
$1-6$ & 7.575E-03 & 3.519E-03 & 1.336E-03 & 4.322E-04 & 1.362E-04 \\
$1-7$ & 5.152E-03 & 2.303E-03 & 8.128E-04 & 2.376E-04 & 6.418E-05 \\
$1-8$ & 6.647E-03 & 6.950E-03 & 7.754E-03 & 8.619E-03 & 7.504E-03 \\
\end{tabular}
\end{table}

\clearpage

\addtocounter {table}{-1}
\begin{table}
\caption{continued.}
\begin{tabular}{cccccc}
\hline
$i-j$ & $15$ Ry & $30$ Ry & $60$ Ry & $120$ Ry & $240$ Ry \\
\hline
$1-42$ & 3.703E-04 & 1.203E-04 & 3.388E-05 & 7.775E-06 & 1.513E-06 \\
$1-43$ & 2.953E-03 & 3.276E-03 & 3.376E-03 & 4.010E-03 & 5.679E-03 \\
$1-44$ & 1.535E-03 & 4.927E-04 & 1.373E-04 & 3.147E-05 & 6.126E-06 \\
$1-45$ & 1.189E-03 & 3.595E-04 & 9.594E-05 & 2.211E-05 & 4.505E-06 \\
$1-46$ & 1.814E-03 & 1.840E-03 & 2.396E-03 & 2.884E-03 & 3.230E-03 \\
$1-47$ & 4.810E-04 & 1.337E-04 & 3.268E-05 & 7.556E-06 & 1.593E-06 \\
$1-48$ & 1.657E-03 & 1.914E-03 & 2.617E-03 & 3.190E-03 & 3.584E-03 \\
\hline
\end{tabular}
\end{table}

\clearpage

\begin{table}
\caption{Fine structure collision strengths from the Al IV level
$1s^{2}2s^{2}2p^{5}3s$ $^{3}$P$^{\rm o}_{2}$ (level 2 in Table
\ref{tab3}) to all the other levels.}
\begin{tabular}{cccccc}
\hline
$i-j$ & $15$ Ry & $30$ Ry & $60$ Ry & $120$ Ry & $240$ Ry \\
\hline
$2-3$ & 1.471E-01 & 1.123E-01 & 1.021E-01 & 1.023E-01 & 1.050E-01 \\
$2-4$ & 4.920E-02 & 4.552E-02 & 4.616E-02 & 4.808E-02 & 4.991E-02 \\
$2-5$ & 9.792E-02 & 4.682E-02 & 2.118E-02 & 1.169E-02 & 8.896E-03 \\
$2-6$ & 1.446E+01 & 1.728E+01 & 2.007E+01 & 2.278E+01 & 2.577E+01 \\
$2-7$ & 4.983E+01 & 6.007E+01 & 7.049E+01 & 8.064E+01 & 9.182E+01 \\
$2-8$ & 1.296E+01 & 1.559E+01 & 1.828E+01 & 2.091E+01 & 2.379E+01 \\
$2-9$ & 1.728E+00 & 2.079E+00 & 2.441E+00 & 2.794E+00 & 3.181E+00 \\
\end{tabular}
\end{table}

\clearpage

\addtocounter {table}{-1}
\begin{table}
\caption{continued.}
\begin{tabular}{cccccc}
\hline
$i-j$ & $15$ Ry & $30$ Ry & $60$ Ry & $120$ Ry & $240$ Ry \\
\hline
$2-42$ & 1.075E-02 & 1.297E-02 & 1.364E-02 & 1.539E-02 & 1.726E-02 \\
$2-43$ & 2.972E-02 & 3.580E-02 & 3.770E-02 & 4.299E-02 & 4.792E-02 \\
$2-44$ & 3.669E-02 & 4.379E-02 & 4.628E-02 & 5.350E-02 & 5.863E-02 \\
$2-45$ & 9.894E-02 & 1.191E-01 & 1.266E-01 & 1.473E-01 & 1.596E-01 \\
$2-46$ & 3.107E-02 & 3.626E-02 & 3.820E-02 & 4.481E-02 & 4.867E-02 \\
$2-47$ & 1.757E-02 & 2.021E-02 & 2.134E-02 & 2.524E-02 & 2.725E-02 \\
$2-48$ & 4.974E-02 & 5.899E-02 & 6.300E-02 & 7.503E-02 & 8.019E-02 \\
\hline
\end{tabular}
\end{table}

\clearpage

\begin{table}
\caption{Fine structure collision strengths from the Al IV level
$1s^{2}2s^{2}2p^{5}3s$ $^{3}$P$^{\rm o}_{1}$ (level 3 in Table
\ref{tab3}) to all the other levels.}
\begin{tabular}{cccccc}
\hline
$i-j$ & $15$ Ry & $30$ Ry & $60$ Ry & $120$ Ry & $240$ Ry \\
\hline
$3-4$ & 3.685E-02 & 1.502E-02 & 5.086E-03 & 1.496E-03 & 3.916E-04 \\
$3-5$ & 6.059E-02 & 3.646E-02 & 2.580E-02 & 2.268E-02 & 2.228E-02 \\
$3-6$ & 6.336E+00 & 7.556E+00 & 8.754E+00 & 9.914E+00 & 1.120E+01 \\
$3-7$ & 5.507E-03 & 1.423E-03 & 3.950E-04 & 2.524E-04 & 1.709E-04 \\
$3-8$ & 2.254E+01 & 2.713E+01 & 3.177E+01 & 3.629E+01 & 4.126E+01 \\
$3-9$ & 1.238E+01 & 1.490E+01 & 1.747E+01 & 1.996E+01 & 2.270E+01 \\
$3-10$ & 5.194E+00 & 6.258E+00 & 7.356E+00 & 8.431E+00 & 9.611E+00 \\
\end{tabular}
\end{table}

\clearpage

\addtocounter {table}{-1}
\begin{table}
\caption{continued.}
\begin{tabular}{cccccc}
\hline
$i-j$ & $15$ Ry & $30$ Ry & $60$ Ry & $120$ Ry & $240$ Ry \\
\hline
$3-42$ & 9.618E-05 & 2.584E-05 & 4.255E-06 & 5.883E-07 & 8.559E-08 \\
$3-43$ & 3.608E-03 & 4.123E-03 & 4.299E-03 & 4.578E-03 & 5.022E-03 \\
$3-44$ & 1.657E-02 & 1.990E-02 & 2.070E-02 & 2.318E-02 & 2.603E-02 \\
$3-45$ & 9.090E-04 & 2.480E-04 & 4.489E-05 & 9.713E-06 & 4.161E-06 \\
$3-46$ & 3.619E-02 & 4.377E-02 & 4.594E-02 & 5.273E-02 & 5.793E-02 \\
$3-47$ & 2.989E-02 & 3.612E-02 & 3.806E-02 & 4.427E-02 & 4.836E-02 \\
$3-48$ & 1.491E-02 & 1.750E-02 & 1.851E-02 & 2.194E-02 & 2.326E-02 \\
\end{tabular}
\end{table}

\clearpage

\begin{table}
\caption{Fine structure collision strengths from the Al IV level
$1s^{2}2s^{2}2p^{5}3s$ $^{3}$P$^{\rm o}_{0}$ (level 4 in Table
\ref{tab3}).}
\begin{tabular}{cccccc}
\hline
$i-j$ & $15$ Ry & $30$ Ry & $60$ Ry & $120$ Ry & $240$ Ry \\
\hline
$4-5$ & 1.733E-02 & 7.144E-03 & 2.359E-03 & 6.720E-04 & 1.699E-04 \\
$4-6$ & 1.864E+00 & 2.219E+00 & 2.566E+00 & 2.902E+00 & 3.275E+00 \\
$4-7$ & 1.176E-04 & 1.330E-04 & 1.410E-04 & 1.307E-04 & 9.220E-05 \\
$4-8$ & 7.840E-04 & 1.847E-04 & 2.633E-05 & 4.108E-06 & 9.574E-07 \\
$4-9$ & 7.347E+00 & 8.834E+00 & 1.034E+01 & 1.180E+01 & 1.340E+01 \\
$4-10$ & 3.378E-04 & 7.343E-05 & 7.813E-06 & 5.985E-07 & 1.022E-07 \\
$4-11$ & 6.675E+00 & 8.045E+00 & 9.453E+00 & 1.082E+01 & 1.232E+01 \\
\end{tabular}
\end{table}

\clearpage

\addtocounter {table}{-1}
\begin{table}
\caption{continued.}
\begin{tabular}{cccccc}
\hline
$i-j$ & $15$ Ry & $30$ Ry & $60$ Ry & $120$ Ry & $240$ Ry \\
\hline
$4-42$ & 2.354E-04 & 2.602E-04 & 2.611E-04 & 2.310E-04 & 1.677E-04 \\
$4-43$ & 1.125E-05 & 3.122E-06 & 5.488E-07 & 8.427E-08 & 1.420E-08 \\
$4-44$ & 1.829E-03 & 2.230E-03 & 2.298E-03 & 2.352E-03 & 2.749E-03 \\
$4-45$ & 3.389E-06 & 3.305E-06 & 2.948E-06 & 2.479E-06 & 1.987E-06 \\
$4-46$ & 1.405E-05 & 3.994E-06 & 7.728E-07 & 1.332E-07 & 2.453E-08 \\
$4-47$ & 1.513E-03 & 1.852E-03 & 1.920E-03 & 2.053E-03 & 2.370E-03 \\
$4-48$ & 1.334E-05 & 3.537E-06 & 5.779E-07 & 7.899E-08 & 1.138E-08 \\
\end{tabular}
\end{table}

\clearpage

\begin{table}
\caption{Fine structure collision strengths from the Al IV level
$1s^{2}2s^{2}2p^{5}3s$ $^{1}$P$^{\rm o}_{1}$ (level 5 in Table
\ref{tab3}).}
\begin{tabular}{cccccc}
\hline
$i-j$ & $15$ Ry & $30$ Ry & $60$ Ry & $120$ Ry & $240$ Ry \\
\hline
$5-6$ & 1.993E-01 & 2.338E-01 & 2.684E-01 & 3.020E-01 & 3.396E-01 \\
$5-7$ & 3.479E-03 & 7.600E-04 & 9.430E-05 & 2.344E-05 & 1.419E-05 \\
$5-8$ & 5.053E-01 & 6.018E-01 & 6.999E-01 & 7.969E-01 & 9.050E-01 \\
$5-9$ & 9.052E-02 & 1.055E-01 & 1.223E-01 & 1.394E-01 & 1.585E-01 \\
$5-10$ & 1.741E+01 & 2.092E+01 & 2.445E+01 & 2.790E+01 & 3.171E+01 \\
$5-11$ & 1.163E+01 & 1.397E+01 & 1.635E+01 & 1.866E+01 & 2.120E+01 \\
$5-12$ & 1.939E+01 & 2.332E+01 & 2.732E+01 & 3.121E+01 & 3.550E+01 \\
\end{tabular}
\end{table}

\clearpage

\addtocounter {table}{-1}
\begin{table}
\caption{continued.}
\begin{tabular}{cccccc}
\hline
$i-j$ & $15$ Ry & $30$ Ry & $60$ Ry & $120$ Ry & $240$ Ry \\
\hline
$5-42$ & 6.740E-05 & 1.816E-05 & 3.060E-06 & 4.349E-07 & 6.580E-08 \\
$5-43$ & 3.529E-04 & 2.163E-04 & 1.906E-04 & 1.766E-04 & 1.364E-04 \\
$5-44$ & 2.091E-03 & 2.133E-03 & 2.045E-03 & 2.380E-03 & 2.403E-03 \\
$5-45$ & 5.906E-04 & 1.583E-04 & 2.618E-05 & 3.947E-06 & 8.148E-07 \\
$5-46$ & 1.453E-02 & 1.720E-02 & 1.719E-02 & 1.939E-02 & 2.060E-02 \\
$5-47$ & 1.151E-02 & 1.353E-02 & 1.345E-02 & 1.534E-02 & 1.612E-02 \\
$5-48$ & 2.208E-02 & 2.647E-02 & 2.661E-02 & 2.979E-02 & 3.206E-02 \\
\hline
\end{tabular}
\end{table}

\end{document}